# The *Helicobacter pylori* AI-Clinician: Harnessing Artificial Intelligence to Personalize *H. pylori* Treatment Recommendations

**Kyle Higgins**[1†], **Olga P. Nyssen**[2†], **Joshua Southern**[3], **Ivan Laponogov**[1], **AIDA CONSORTIUM**[4], **Dennis Veselkov**[1,3], **Javier P. Gisbert**[2*], **Tania Fleitas Kanonnikoff**[5*] and **Kirill Veselkov**[1,6*].

[1] Division of Cancer, Department of Surgery and Cancer, Faculty of Medicine, Imperial College London, London, UK.
[2] Gastroenterology Unit, Hospital Universitario de La Princesa, Instituto de Investigación Sanitaria Princesa (IIS-Princesa), Universidad Autónoma de Madrid (UAM), Centro de Investigación Biomédica en Red de Enfermedades Hepáticas y Digestivas (CIBEREHD), Madrid, Spain
[3] Department of Computing, Imperial College London, U.K.
[4] The full list of members and affiliations of The AIDA Consortium is provided in the Supplementary Information.
[5] Instituto Investigación Sanitaria INCLIVA (INCLIVA), CIBERONC, Medical Oncology Department, Hospital Clínico Universitario de Valencia, Universitat de Valencia, Valencia, Spain
[6] Department of Environmental Health Sciences, Yale University, New Haven, CT, USA.
[†] Shared first authorship
[*] To whom correspondence should be addressed: kirill.veselkov04@imperial.ac.uk, tfleitas@incliva.es, javier.p.gisbert@gmail.com

## Abstract

*Helicobacter pylori (H. pylori)* is the most common carcinogenic pathogen worldwide. Infecting roughly 1 in 2 individuals globally, it is the leading cause of peptic ulcer disease, chronic gastritis, and gastric cancer. To investigate whether personalized treatments would be optimal for patients suffering from infection, we developed the *H. pylori* AI-clinician recommendation system. This system was trained on data from tens of thousands of *H. pylori*-infected patients from *Hp*-EuReg, orders of magnitude greater than those experienced by a single real-world clinician. We first used a simulated dataset and demonstrated the ability of our AI Clinician method to identify patient subgroups that would benefit from differential optimal treatments. Next, we trained the AI Clinician on *Hp*-EuReg, demonstrating on average the AI Clinician reproduces known quality estimates of treatment decision making, for example bismuth and quadruple therapies out-performing triple, with longer durations and higher dose proton pump inhibitor (PPI) showing higher quality estimation on average. Next, we demonstrated that treatment was optimized by recommended personalized therapies in patient subsets, where 65% of patients were recommended a bismuth therapy of either metronidazole, tetracycline, and bismuth salts with PPI, or bismuth quadruple therapy with clarithromycin, amoxicillin, and bismuth salts with PPI, and 15% of patients recommended a quadruple non-bismuth therapy of clarithromycin, amoxicillin, and metronidazole with PPI. Finally, we determined trends in patient variables driving the personalized recommendations using random forest modelling. With around half of the world likely to experience *H. pylori* infection at some point in their lives, the identification of personalized optimal treatments will be crucial in both gastric cancer prevention and quality of life improvements for countless individuals worldwide.

**Background**

*Helicobacter pylori (H. pylori)* is a Gram-negative S-shaped bacteria which has adapted to colonize the niche of the deep gastric mucous layer in the human stomach.[1] *H. pylori* infection in the gastric mucosa leads to a diverse inflammatory response in local epithelial cells, resulting in chronic active gastritis. (**Figure 1A**) Despite producing antibodies to *H. pylori* antigens, this immune response is generally incapable of eradicating the bacteria. Over decades, this inflammation has been thought to lead to a variety of conditions, most notably peptic ulcer disease (PUD) and gastric cancer. Of the nearly 4 billion people infected by *H. pylori*, approximately 10% will develop PUD within a decade of infection, meaning roughly 780 million worldwide will be afflicted by this condition.[2] (**Figure 1B**) Peptic ulcers result from damage to the lining of the stomach and may lead to complications such as internal bleeding and perforations, with a high mortality rate in such cases. *H. pylori* eradication has shown promising results in the treatment of PUD, achieving not only ulcer healing but also preventing its recurrence.

Around 90% of gastric cancer cases are due to *H. pylori* infection.[3] It is estimated that gastric cancer makes up 37% of chronic infection-induced cancers, making *H. pylori* the most frequently carcinogenic pathogen.[4] Gastric cancer is thought to develop after years of inflammation-induced gastric atrophy, wherein achlorhydria drives the development of an abnormal microbiome, further driving the transformation of gastric epithelial cells to an oncogenic state (a hypothesis termed the 'Correa cascade').[5] Indeed, a 'point of no return' has been observed with regards to *H. pylori* infection in patients developing gastric cancer, past

**(A)** *Helicobacter pylori* infection

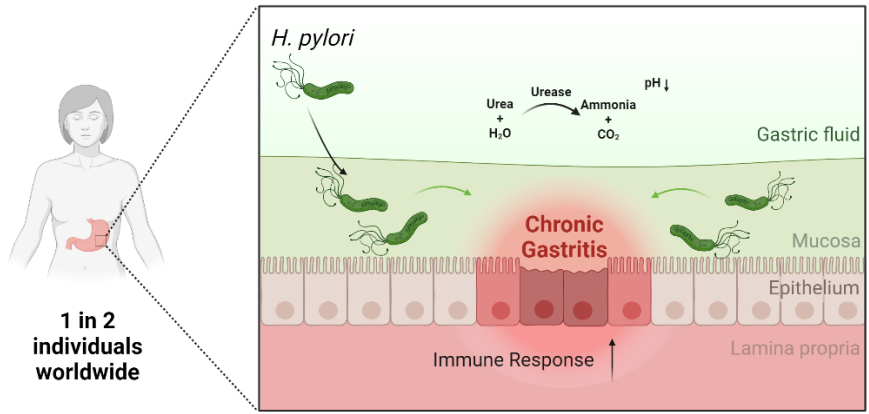

**(B)** Pathology

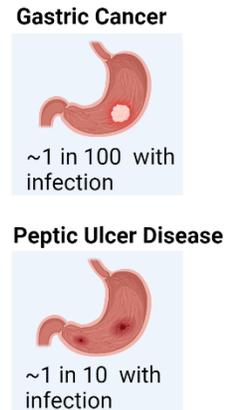

**(C)** Hp-EuReg

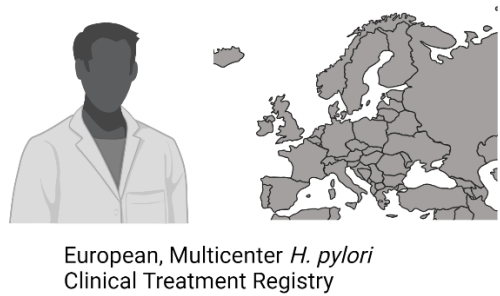

**(D)** Data Structure

| Patient Metadata | Treatment Strategy (Clinician) | Outcome |
|---|---|---|
| Female, Spain, 35, .. | C+A+M+PPI, 14 Day | Eradication |
| Male, Spain, 62, .. | A+M+PPI, 10 Day | Eradication |
| Male, Russia, 65, .. | C+A+M+PPI, 14 Day | Failure |
| Female, UK, 54, .. | A+T+PPI, 10 Day | Failure |
| Male, Italy, 28, .. | Pylera, 10 Day | Eradication |
| ... | ... | ... |

**(E)** AI-Clinician

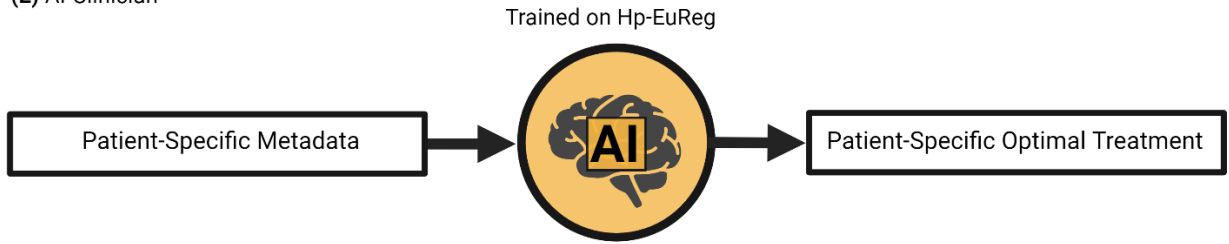

**Figure 1 Overview.** (**A**) *Helicobacter pylori* (*H. pylori*) infects the stomach of around one in two individuals worldwide. It does do by infiltrating the gastric mucosa, aided by a highly motile flagellum. The infection is characterized by an overall increase in the acidity of the gastric fluid, onset by urease production, and increased inflammation in the gastric epithelia, often spanning decades before diagnosis. (**B**) *H. pylori*-induced pathology most commonly includes gastric cancer (at least one per one hundred infected individuals) and peptic ulcer disease (around one in ten infected individuals). (**C**) The Hp-EuReg project is an international, multicenter prospective registry collecting *H. pylori* treatment management strategies across Europe and including to date over 75,000 patient records. (**D**) The data included in this registry includes patient metadata, treatment strategy employed by the clinician, and result of this treatment, in terms of eradication. (**E**) The *H. pylori* AI-clinician is trained on the Hp-EuReg dataset and designed to provide patient-specific optimal treatment recommendations for *H. pylori* eradication.

which *H. pylori* eradication is insufficient to interrupt the inflammatory cascade leading to oncogenesis. Despite advancements in treatment such as chemotherapy and surgery, gastric cancer results in a poor prognosis compared to other cancer types, especially in advanced stages.[6]

The European Registry on *Helicobacter pylori* management (Hp-EuReg) was established to combat the high social and health burden of *H. pylori* infection across Europe. It was noted at the time that consensus and clinical guidelines were established for *H. pylori* treatment, but that no data existed cataloging the implementation of these recommendations.[7] This project took the form of an international and multicenter prospective non-interventional registry documenting the real clinical practice by European gastroenterologists of *H. pylori* management in the majority of countries across Europe. (**Figure 1C**) Patient data documented include several demographics categories (i.e., country, sex, age), pre-existing gastrointestinal symptoms, treatment indication, previous eradication attempts, and compliance. (**Figure 1D**) Crucially, this registry documents treatment chosen, duration of treatment, proton pump inhibitor (PPI) dosage and eradication outcome. To date, this registry has been used in over 40 published studies.[8] (For a full list of publications, see [9].) The most common uses for this dataset have been to assess treatment effectiveness, especially in a country or region-specific context. However, most of these studies have mainly relied on traditional statistical methods rather than advanced methods employing machine learning (ML), artificial intelligence (AI), with one notable exception[10].

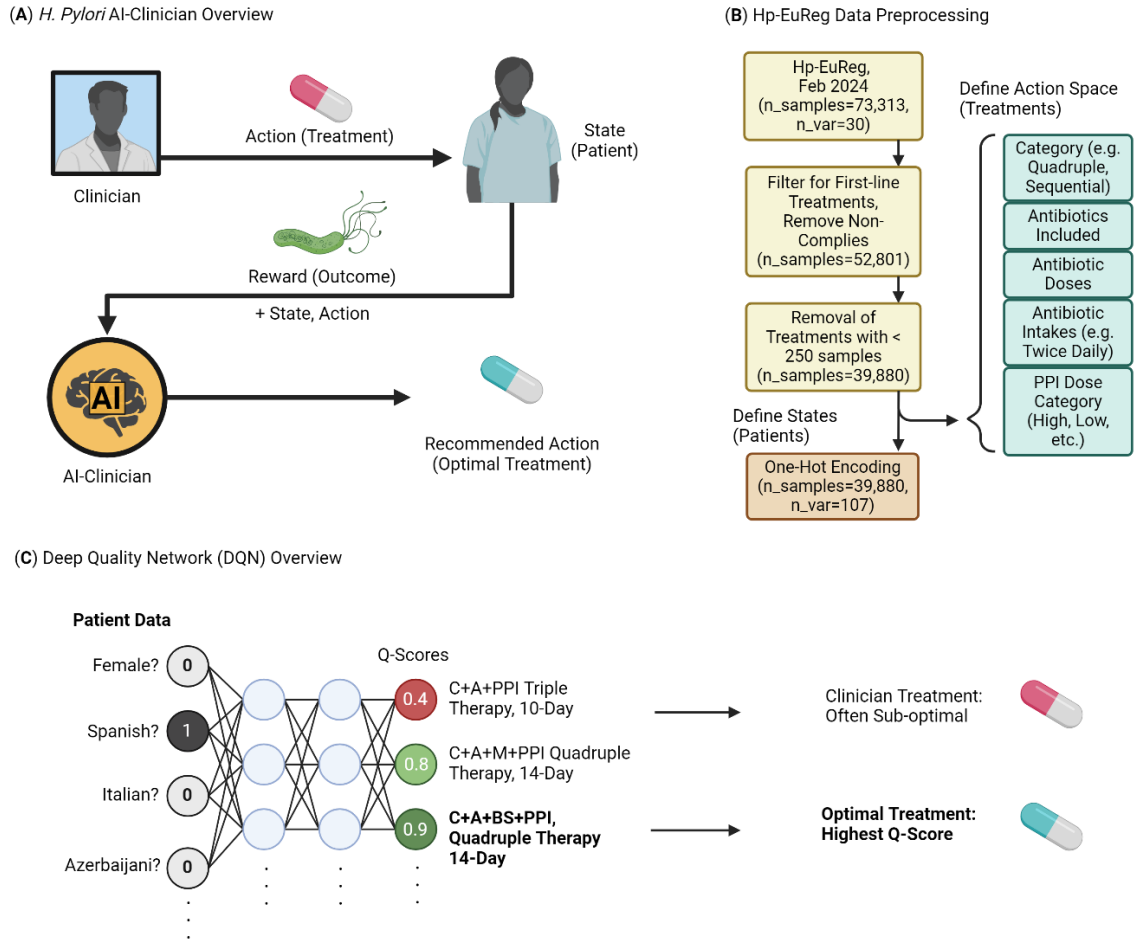

**Figure 2 Methods Overview.** (**A**) The *H. Pylori* AI-Clinician is a recommendation system trained via reinforcement learning (RL) which learns from the result of clinical decision-making. Mathematically, the AI learns the quality of state, action pairs by observing the reward obtained by each. In practice, states are represented by patient data, and actions are represented by clinical treatment decisions. Reward is measured by the success or failure of treatment. Once trained, the AI-Clinician returns an optimal treatment for individual patients. (**B**) Hp-EuReg data, including 73,313 patient records and 30 pre-treatment patient variable categories, is preprocessed prior to model training. Only first-line treatments and patients who complied with treatment are considered. (52,801 samples) Patient/Treatment pairs with less than 250 samples for a given treatment are removed to ensure sufficient training data. Actions (treatments) are encoded to include treatment category, antibiotics included, doses of antibiotics, and PPI Dose Category (including Low, Standard, High, and Other). State (patient) variables are one-hot encoded, resulting in 39,880 samples containing 107 patient variables total. (**C**) Deep Quality Network (DQN) analysis is implemented to train our recommendation system to identify optimal treatments. One-hot encoded patient data is fed into the network, which is followed by feeding into two hidden layers, and finally an output layer which represents the quality of implementing a given treatment for all possible treatments. The treatment with the highest quality (Q-Score) is the optimal treatment for a given patient. The network is trained via gradient descent at select optimization intervals (every 100 patients, optimized on a memory of 10000 patients). During testing, patient state information is fed into the model to receive optimal treatment recommendations, but no further optimizations are performed.

The most frequently used treatments in this registry include the administration of triple and quadruple (either bismuth and non-bismuth based) antibiotics regimens. Treatment durations predominately include 7, 10, and 14-day prescriptions. Components of these treatments include a combination of at least two antibiotics and a PPI in order to raise stomach pH and bismuth for its bacteriostatic effect. Standard triple therapies, most often consisting of two antibiotics (amoxicillin and clarithromycin), and a PPI, were a great advance in the treatment of *H. pylori* in the 1990s, leading to its once adoption as the treatment gold standard.[11] However, increasing clarithromycin resistance (up to 23% observed in Hp-EuReg)[8] and other factors have led to the development of additional therapies such as quadruple regimens. Certain formulations of quadruple therapies quickly demonstrated >90% eradication rate (now considered the threshold for an optimal *H. pylori* regimen[12]), leading to their adoption as the current recommendation standard[13]. In their traditional formulation, they combine a nitroimidazole (such as metronidazole or tinidazole) with a PPI and antibiotics amoxicillin, clarithromycin. However due in part to an antibiotic resistance, an alternative bismuth quadruple regimen has been widely adopted (including tetracycline and metronidazole), to great effect in first-line treatments [14,15]. Bismuth has been included for its bactericidal effect, rendering bismuth quadruple therapy unaffected by clarithromycin and metronidazole resistance.[16] Sequential therapies were also developed in part to overcome limitations posed by triple therapies, and consist of a two-part treatment period, first using a PPI and amoxicillin, followed by a PPI, clarithromycin, and either tinidazole or metronidazole[17]. However, sequential therapies have been administered to variable effect, with eradication rates varying from <80%-90%, largely dependent on region.[18-20] Finally, bismuth single capsule therapies such as Pylera® replace multi-drug regimens with a single pill containing bismuth, metronidazole, and tetracycline, combined with a PPI.[21] Bismuth single capsule therapy are a relatively new treatment, with Pylera® first approved by the FDA in 2006 and currently only approved in Europe in a subset of countries.[22] Early studies show the eradication rate varying around 80% to around 95%[23,24], and a recent meta-analysis report an effective eradication (90%) not only in first-line but also in rescue therapy and in those patients with clarithromycin- or metronidazole-resistant strains, and in those previously treated with clarithromycin.[25] Though further study is needed as implementation is increased in diverse populations.

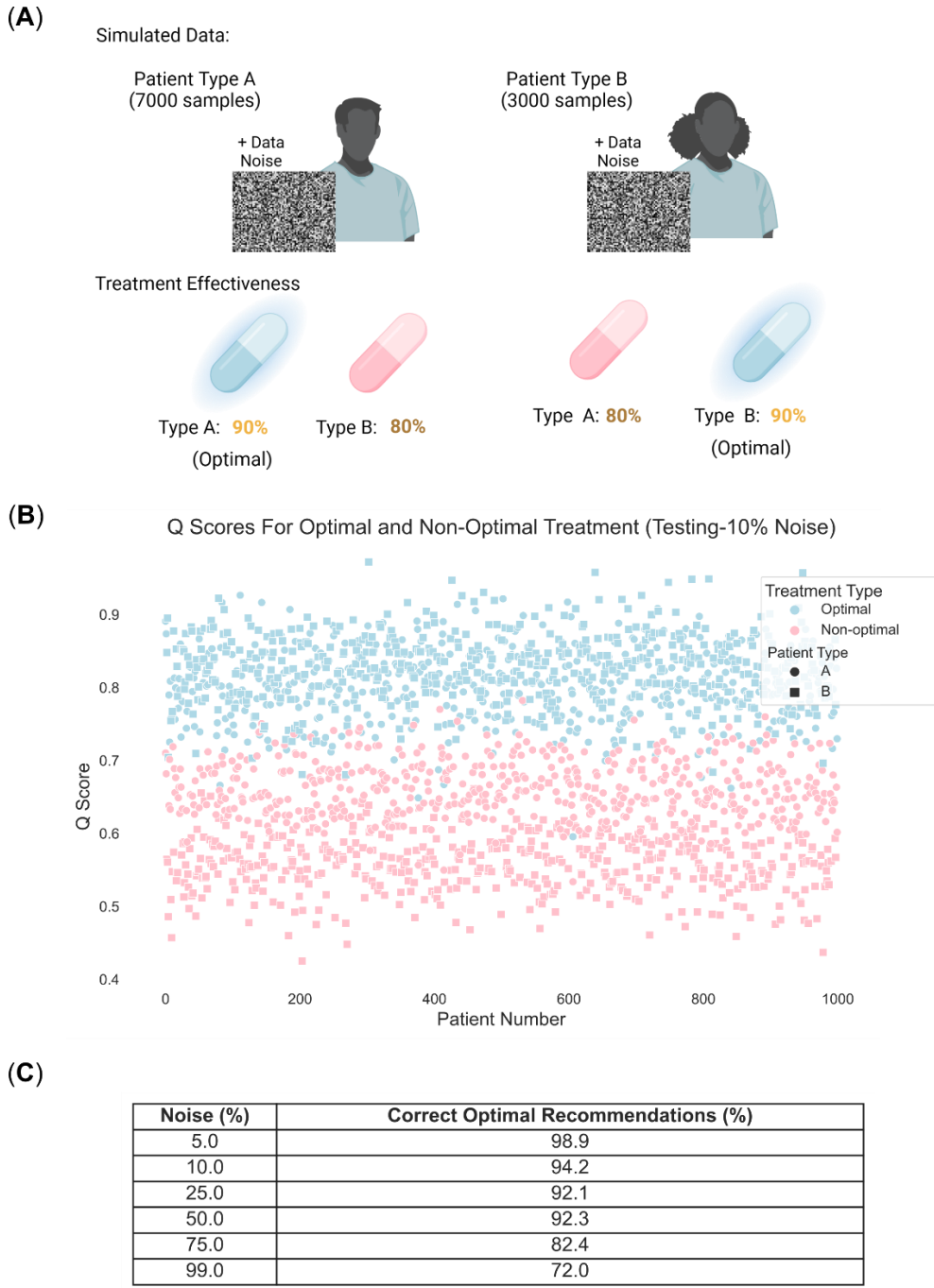

**Figure 3 Performance on Simulated Data.** (**A**) A theoretical dataset was generated to demonstrate the sensitivity of the AI Clinician in detecting optimal treatment relative to a heterogenous patient population, where drug A is 90% effective in population A and drug B is 80%, and vice versa for population B. The theoretical dataset consisted of 10K patients with 100 binary variable features, and distributed into a randomized and balanced 90-10 training-testing split. Reward is assigned as +1 if successfully treated and -1 if unsuccessfully treated. To test the model's performance in the presence of unbalanced classes, 7000 patients of population A are generated and 3000 of populated B. Noise was added to the dataset as random introductions of 0's and 1's at varying levels to simulate increasingly noisy real-world data. (**B**) Q Scores for recommendations of both the optimal and non-optimal treatment (two per patient) were plotted during the testing phase, with recommendations of the drug optimal for that simulated patient highlighted in green and patient type distinguished by shape. Overall, the AI clinician correctly ascribes optimal treatments for both patient types with higher Q scores than non-optimal, with variation due to noise. (**C**) Repeating analysis over many different noise levels showed that the majority of patients are correctly recommended the optimal drug even with increasing noise, demonstrating the ability of the model to detect heterogeneity in patient populations even in highly noisy data.

A further consideration for treatment recommendation is the presence of allergies to penicillin-like medications, such as amoxicillin. Around 1-5% of patients globally have documented penicillin allergies[26], though higher percentages suggested when self-reporting.[27] The presence of this allergy necessitates use of therapies without amoxicillin, such as levofloxacin-based regimens or a tetracycline, metronidazole, and bismuth salts regimens combined with a PPI such as those found in bismuth single capsule therapies.

Recently, AI and ML have revolutionized several fields including medicine in part due to their propensity for patterns in large quantities of data. Several AI approaches have had particular success. Supervised learning involves the training of models on labeled data to either classify or make predictions about data from other sources. Unsupervised learning on the other hand operates on unlabeled data, with the purpose of finding unknown patterns in the data. Common examples of this technique include clustering and dimensionality reduction. Finally, reinforcement learning (RL) is a paradigm designed for the purpose of training a decision-making apparatus by rewarding desirable decisions and penalizing undesirable ones. It has found particular success in the fields of robotics, video gaming, and various optimization problems. However, its utility has rarely been applied in the field of medicine.

We developed the *H. pylori* AI-Clinician which applies RL to determine patient-specific first-line treatment recommendations and determine one size fits all for optimal treatment strategy. (**Figure 1E**) This method applies RL, which in this case learns iteratively which actions to take (termed policy) to maximize reward in the context of a given state. RL is well-suited to datasets such as the case of the Hp-EuReg with many interacting variables (on the order of a thousand when one-hot encoded) as it is sensitive to small differences in rewards, and able to detect subtle factors which affect outcome in the state, and therefore patient outcomes.[28]

**Method Development**

The *H. pylori* AI-clinician was developed in order to predict optimal treatment outcomes on a patient-specific basis while learning from real-world clinical decisions. It was developed using RL, which is a ML approach by which an agent learns by observing the reward it receives from taking actions in a trial-and-error manner. However, since our agent (the AI-clinician) was being applied to real-world patients, it is both dangerous and unethical to implement trial-and-error learning in early stages. Therefore, an alternative approach was implemented which trains the agent using clinical actions (treatments) and their observed reward (success/failure) rather than implementing the actions of the agent directly. (**Figure 2A**)

The specific RL method developed was termed independent-state Deep Q-Network Learning. (isDQN) It is an adaptation of Deep Q-Network (DQN) Learning, which is one of the most widely applied methods in RL. [29-32] Q-learning is a method by which the quality of state-action pairs is learned over time, where the state is some set of environmental variables and the actions are the set of all possible actions an agent may take. Usually, the action taken will be that with the highest Q-score in a given state context. DQN is termed 'deep' in that it utilizes a neural network of several layers to achieve its learning. Reward is observed for each action in the context of a given state, and this reward is in turn used to alter weights in the neural network during some optimization step to improve the estimation of Q-scores in the future, therefore improving decision-making. Optimizations are traditionally some functions of both immediate reward from action taken, and the quality of the state the action puts the environment in in the subsequent moment (for example, increasing score in a video game immediately, and also putting the agent in a position to further increase score in future moves.)

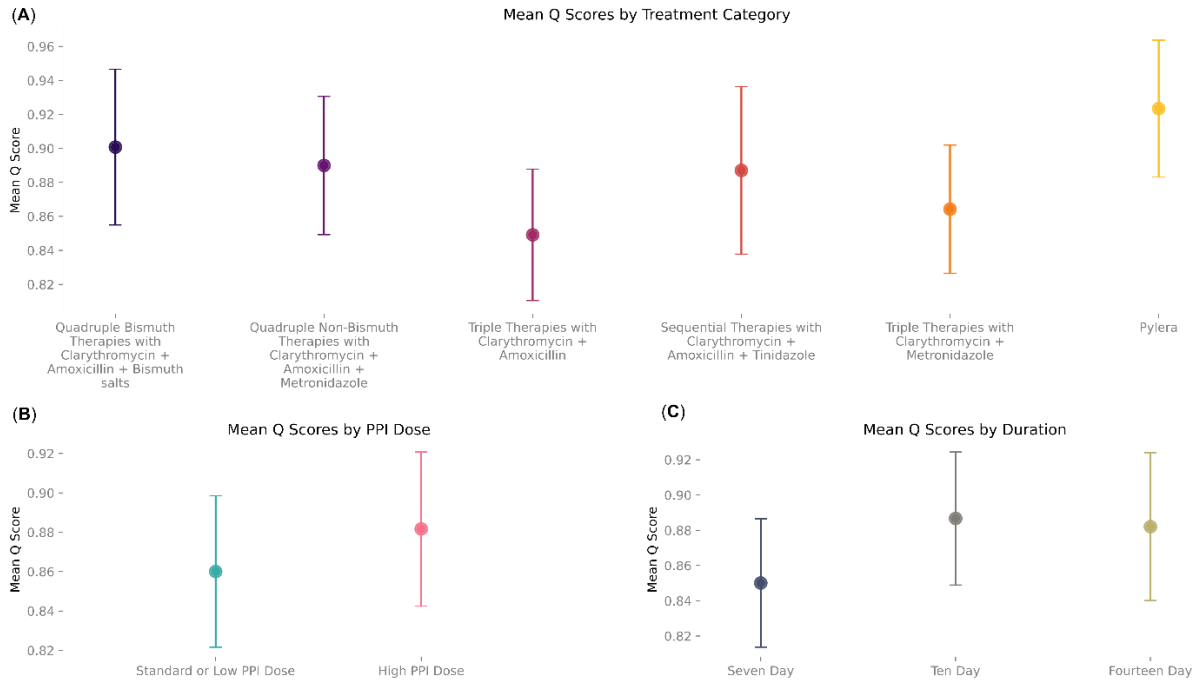

**Figure 4** *H. pylori* **AI-Clinician Training and Performance on Real-World Data.** (**A**) The mean Q scores of each treatment category in the testing phase are compared, demonstrating the AI Clinician's preference for treatments on average. Pylera has the highest overall Q score (mean=0.92, SD=0.04), followed by quadruple bismuth therapies (mean=.90, SD=0.05), quadruple non-bismuth therapies (mean=0.89, SD=0.04), and sequential therapies (mean=0.89, SD=0.05). triple therapies have the lowest Q score on average (mean=0.86,0.85; SD=0.04,0.05 for clarithromycin + metronidazole and clarithromycin + amoxicillin Therapies, respectively). All treatments include the prescription of a PPI. (**B**) Mean Q Scores by PPI Dose demonstrate High PPI dose (mean=0.88, SD=0.04) has a higher average Q score than standard or low dose PPI (mean=0.86, SD=0.04). (**C**) Mean Q scores by duration of treatment also demonstrate that 10 and 14 day durations (mean=0.89,0.88; SD=0.04,0.04, respectively) out-perform 7 day (mean=0.85,SD=0.04), though result in similar Q scores compared to one another.

Our method, isDQN differs from traditional RL methods in that there is no concept of a 'subsequent state' for patient data as this data remains static before and after treatment, with eradication fully represented in the reward. Mathematically, optimization is traditionally performed using a loss function which aims to represent and minimize what 'cost' was accrued by the current failures in decision-making at a given step. This loss function is traditionally calculated using the Q-score of the current state subtracted from the immediate reward and maximum expected quality of the subsequent state. In our method, the quality of the subsequent state is not considered, effectively eliminating it from the loss function. (In mathematically rigorous terms, the trajectory is always taken to be in its final state, which is a special case described in the original formulation of DQN analysis detailed by Mnih et al.[33]

The representation of patient information is as follows. States were represented by the one-hot encoded patient variables represented in the Hp-EuReg dataset. Actions were also one-hot encoded as described by clinical treatment decisions including antibiotic/PPI combination, antibiotic dose, PPI dose category (see *Methods* for values), and duration (for example, clarythromycin + amoxicillin + metronidazole + clarithromycin dose = 500 mg, + intakes twice daily + ... + high dose PPI, 14-day duration) and encoded numerically. Only treatments with at least 500 examples in the dataset were included in the action space. (**Figure 2B**) Note that PPI dose category was divided into categories of 'Standard or Low Dose' and 'High Dose' to reduce the number of treatment category divisions. Reward was represented by the clinical outcome of the treatment, therefore a value of +1 if eradication was achieved and -1 if the eradication was a failure. The model was trained using batches of patients, where a mathematical operation termed gradient descent reweighs the neural network at each optimization step using a mean squared error (MSE) loss function. Over time, this agent learns the quality of each treatment for a given patient, determining patient-specific optimal treatments. (**Figure 2C**)

**Results**

To demonstrate the effectiveness of our model in detecting heterogeneous optimal treatment choices, we generated a simulated dataset of 10,000 patients with 100 binary features. To model an asymmetric dataset (unbalanced classes) 7000 patients were assigned one random combination of binary variables, while another group of 3000 was assigned a different random combination of variables. (**Figure 3A**) Noise was added to the dataset at various levels to obscure the structure of variables determining outcome, similar to real-world datasets. Two simulated treatments were used in the dataset, where treatment A was given 90% success in group A of patients, and treatment B was given 80% success in group A. Likewise, treatment B was given 90% success in group B whereas it was only given 80% in group A.

Data was generated and split into a balanced and randomized 90-10 training-testing distribution. Here we consider a 'correct' optimal drug recommendation to be one where the drug with 90% effectiveness in the patients' group of origin is recommended (For example, drug A to a patient from group A). During the testing phase, Q scores for both treatments (optimal and non-optimal) were examined, demonstrating highly accurate optimal drug recommendations. (**Figure 3B**) Noise levels of 5, 10, 25, 50, 75, and 99% were tested to evaluate the model's performance in the presence of noisy data. The percentage of patients correctly identified in each noise level was 98.9% (5% noise), 94.2% (10% noise), 92.1% (25% noise), 92.3% (50% noise), 82.4% (75%), 72.0% (99% noise) demonstrating that the model is capable of detecting patient heterogeneity in optimal treatment even in the presence of highly noisy data. (**Figure 3C**)

A second, more complex simulated analysis was performed for hyperparameter tuning. In this case, 10 unique simulated patient 'types' were generated, with a unique random pattern of 70 one-hot encoded patient variables, for 50,000 simulated patients total. In each simulated patient 'type' a single optimal treatment out of 10 possibilities is given 90% effectiveness, with all other treatments providing 70% effectiveness. An additional 5% noise is added to the data to simulate real world variability. A hyperparameter grid testing multiple combinations of batch size, steps per optimization, deque size, and learning rate is sampled to include every combination, and a model trained using a random sample of 80% of the data. As a metric of success, a score is generated between 0 and 1 which represents the fraction of patients in the testing dataset (remaining 20%) which was correctly assigned the treatment corresponding to 90% success rate in the simulated training group. An optimal combination favoring small batch sizes and a fast learning rate was found to perform best and used for analysis of real-world data. (For a full description, see *Methods*)

Moving to Hp-EuReg, we evaluated the consistency of the AI Clinician with different splits of training data by generating 500 independent models by ten-fold, fifty repeat cross validation. For each repeat, training (model optimization) was performed using a 90% random sample of first-line treatments from the Hp-EuReg dataset, with testing performed on the remaining 10%. AI performance was compared to clinicians by comparing Q-scores of the clinician's action to the AI-recommended action for each patient. In a representative model, mean Q scores in the testing phase were tabulated to quantify the AI Clinician's preference for different treatment categories. All therapies include prescription of a PPI. Overall, Pylera® therapies had the highest average Q score (mean=0.92, SD=0.04), suggesting that over the entire testing dataset it was on average estimated to be the most effective on a diverse population of patients. It was followed by quadruple bismuth therapies with clarithromycin, amoxicillin, and bismuth salts (mean=.90, SD=0.05), quadruple non-bismuth therapies with clarithromycin, amoxicillin, and metronidazole (mean=0.89, SD=0.04), and sequential therapies with clarithromycin, amoxicillin, and tinidazole (mean=0.89, SD=0.05). Triple therapies performed the most poorly, with clarithromycin and metronidazole (mean=0.86, SD=0.04) slightly outperforming clarithromycin and amoxicillin (mean=0.85, SD=0.05). (**Figure 4A**) Average Q scores based on PPI dose demonstrate a preference for high dose PPI on average (mean=0.88, SD=0.04) compared to low or standard doses (mean=0.86, SD=0.04). (**Figure 4B**) For a full description of PPI dose definitions given particular PPIs, see *Methods*. It is worth noting that many patients in the training dataset had a PPI dose which was not specified by the clinician. However, the category of patients with unspecified dose also showed high Q scores (mean=0.89, SD=0.04) suggesting this population was dominated by high PPI doses. Finally, ten (mean=0.89, SD=0.04) and fourteen-day (mean=0.88, SD=0.04) durations were found to have higher Q scores than seven day (mean=0.85, SD=0.04) treatment periods—with ten and fourteen day periods showing similar average Q scores. (**Figure 4C**)

When tabulated over all repeats (50 recommendations per patient in the testing phase), 65.5% of patients were consistently recommended a bismuth therapy consisting of either Pylera® or clarithromycin, amoxicillin, and bismuth salts paired with a PPI by more than half of the models, which we treat as a frequency threshold. Further, 15.5% of patients were recommended a non-bismuth clarithromycin, amoxicillin, and metronidazole with PPI, and 19.0% of patients recommended a variety of therapies, with no single therapy being recommended by more than half of models. Notably no patients were consistently recommended triple or sequential therapies. (**Figure 5A**) We additionally examined the breakdown of patients where unique bismuth therapies are considered separately, where patients need to be recommended either Pylera® or quadruple bismuth therapies with clarithromycin, amoxicillin, and bismuth salts by more than half of models to be considered above frequency threshold. 51.5% were recommended quadruple non-bismuth therapy with clarithromycin, amoxicillin, and

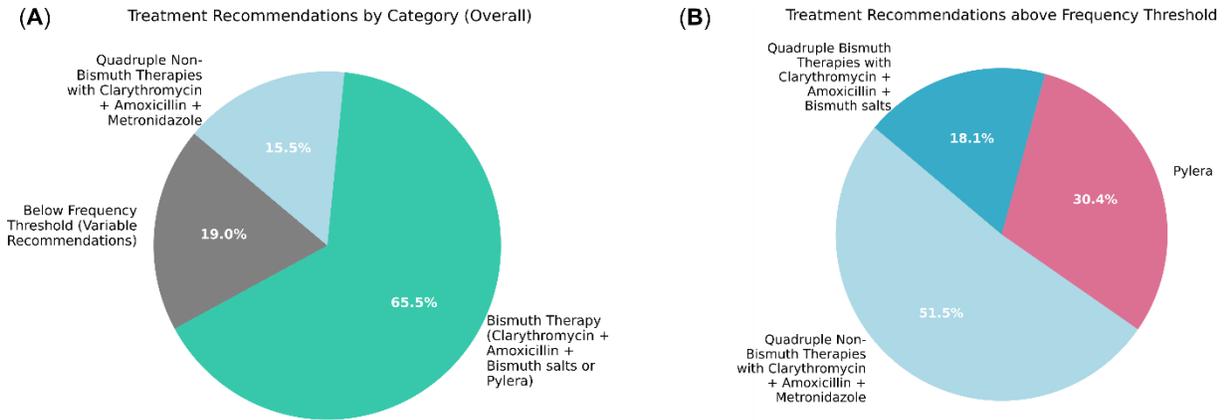

(C)

|  | Pylera | Bismuth Quadruple | Bismuth Salts (Any) | Non-Bismuth Quadruple |
|---|---|---|---|---|
| Highly Associated Variables | Southwest Region | Taking Concurrent Medication (Any) | Eastern Region | Eastern Region |
|  | Not taking Acetylsalicylic Acid | Eastern Region | Taking Concurrent Medication (Any) | Caucasian |
|  | Taking Concurrent Medication (Any) | Not taking probiotics | Not Experiencing Heartburn | Not taking Rebamipid |
| Balanced Accuracy (%) | 84.7 | 76.7 | 73.6 | 92.3 |

**Figure 5 Personalized Recommendations.** (**A**) Recommendations per individual across 50 repeated training-testing cycles on various splits of the data were generated, with mode treatment category tabulated for each patient, with the requirement that it was recommended by more than half of the repeats. On average, 65.5% of patients were recommended a Bismuth Therapy consisting of either Pylera® or clarithromycin, amoxicillin, and bismuth salts (with PPI), 15.5% of patients were recommended non-bismuth quadruple therapy with clarithromycin, amoxicillin, and metronidazole (with PPI), and 19.0% were recommended variable treatments, with no majority recommendation. (**B**) When Pylera® and quadruple therapy with bismuth salts are distinguished, 9.2% of patients are recommended Pylera® more than half of the time and 5.4% are recommended quadruple therapy with bismuth salts, suggesting that there is not a strong preference for which of the two therapies was recommended for the majority of patients which are routinely recommended a bismuth therapy. (**C**) Random Forest models are generated for each of the therapy categories discussed above to discover the relevance of patient variables in determining recommended therapy. Variables are ranked by mean decrease in impurity (MDI) to determine the top three variables most associated with a particular treatment recommendation.

metronidazole, 30.4% were recommended Pylera®, and 18.1% were recommended quadruple bismuth therapies with clarithromycin, amoxicillin, and bismuth salts. (**Figure 5B**)

Finally, in order to see which variables correlated most strongly to treatment recommendation, we perform RF analysis each time for four recommendation groups: Pylera®, bismuth quadruple (with clarithromycin, amoxicillin, and bismuth salts), bismuth salts (any) which could be either Pylera® or bismuth quadruple therapy, and non-bismuth quadruple therapy (with clarithromycin, amoxicillin, and metronidazole). The model is formulated to predict whether a patient will receive a given treatment versus all others, based on patient variables. RF models had a balanced accuracy of prediction of 84.7% for Pylera®, 76.7% for bismuth quadruple therapies, 73.6% for bismuth salts (any), and 92.3% for non-bismuth quadruple therapies. Patient variable importance was ranked by mean decrease in impurity (MDI) to determine highest predictive power for a given treatment recommendation. Overall, being from the southwest region of Europe, not taking acetylsalicylic acid, and taking concurrent medication of any kind were more likely to result in a Pylera® recommendation. Taking concurrent medication, being from an eastern region of Europe and not taking probiotics were more likely result in a bismuth quadruple recommendation. Likewise, being from an eastern region, taking any concurrent medication, and not experiencing heartburn as a symptom were more likely to correspond to a bismuth salts therapy of any sort. Finally, being from the eastern region, Caucasian, and not taking rebamipid were more likely to correspond to a recommendation of non-bismuth quadruple therapy. (**Figure 5C**)

## Discussion

The *H. pylori* AI-clinician was developed to investigate whether optimal treatment should include prescribing subsets of patients' different therapies depending on their clinical variables. We found that this was the case. Overall, we found that Q scoring in individual models was in line with current trends in treatment recommendations [14,34], demonstrating the reliability of the AI Clinician method.

Pylera® and quadruple therapies with clarithromycin, amoxicillin, and bismuth salts or metronidazole, and sequential therapies showed the highest quality estimate by the AI clinician, in that order (and above triple therapies containing clarithromycin and amoxicillin or metronidazole). We also found that higher dose PPIs performed better than low and standard dose on average, suggesting most patients would benefit from a higher dose. Finally, while 10- and 14-day durations out-performed 7 day, they performed quite similarly in terms of quality estimate to one another, likely driven by Pylera®'s increased effectiveness and 10-day formulation.

We found that 65.5% of patients were recommended a bismuth therapy by the majority of AI Clinician models trained on differing splits of data; while 15.5% were consistently recommended a non-bismuth quadruple therapy with clarithromycin, amoxicillin, and metronidazole. Overall, RF modelling was able to achieve a high balanced accuracy for the latter therapy, indicating that variables including presence in an eastern region, being Caucasian, and not taking rebamipid were highly indicative of a patient receiving a non-bismuth quadruple therapy recommendation. Pylera® was more likely to be recommended if a patient was from a southwest region and taking concurrent medication, but not acetylsalicylic acid. Non-bismuth quadruple therapies including clarithromycin, amoxicillin, and metronidazole were more likely to be recommended again if taking a concurrent medication, but instead from an eastern region and not taking probiotics.

The correspondence to region in personalized recommendations suggests several possibilities: that strains of *H. pylori* varying by region are driving the trend, that interactions between genetics and treatment are responsible, that trends in lifestyle varying by region are responsible, or a combination of the three. Further investigations to study a higher number of region-specific variables in detail will be needed to determine the explanation. Interestingly, though Pylera® did see the highest average Q score from modelling, it was not the most frequently recommended therapy overall (which was non-bismuth quadruple therapies), suggesting that the quality of Pylera® in patients consistently recommended it is higher than other treatments; which is to say these patients have a higher benefit from their optimal therapy than other groups.

Interestingly, 19.0% of patients were recommended a variety of treatments, with no single category being recommended by more than half of models. In addition, most patients were not recommended the same treatment by every model generated by differing splits of training data. A further shortcoming of this study is that several treatment formulations (for example, the quadruple bismuth therapy including metronidazole, tetracycline, and bismuth subcitrate with PPI-which the single capsule Pylera® is based on) were not present in sufficient numbers in the dataset to be included in training due to the requirement of around 500 samples for stabilization in the network observed in our study. PPI dosage category was also divided into only two categories ('Standard or Low' and High') to reduce the number of therapy type subdivisions due to the limited number of samples of each treatment category, though there are well-documented differences in effect between low and standard PPI doses. These facts signal the need to collect additional data for model training and for the improvement of the sensitivity of the AI Clinician in the future.

Though these results are encouraging in terms of increasing early eradication of *H. pylori,* future work should be focused on treatment of patients for which the damage of long-term infection has already been done. For example, a well-established point of no return for infections exists, past which gastric cancer often develops even after eradication of *H. pylori*. Investigating other data types such as endoscopy images and omics data, especially with advanced methods such as ML and AI will be crucial for determining what specific consequences of infection a patient is experiencing or at risk of experiencing, and what therapeutic strategies may be applied.

With over half the globe experiencing *H. pylori* infection at some point in their lifetime, there is a great need to apply advances in AI to evaluate and improve management and treatment, especially with regards to treatment recommendation standards. This work demonstrates the robustness of current recommendation standards throughout a diverse and heterogeneous population, lending support their broad administration. Further, this work demonstrates a fundamentally novel system for making personalized treatment recommendations based on patient data, opening the door to many potential future applications.

Methods

*Variable Preprocessing*

The Hp-EuReg Dataset was obtained on February 14[th], 2024 and was taken from the AEG-REDCap platform. Data were recorded in an Electronic Case Report Form (e-CRF) using the collaborative research platform REDCap hosted at "Asociación Española de Gastroenterología", a non-profit Scientific and Medical Society focused on Gastroenterology research.[35] The dataset consisted of 73,313 patients and 321 patient variables, including treatment administered and outcome in terms of *H. pylori* eradication. Samples were filtered to include only first-line treatments administered to patients who complied with their full regimen

**Algorithm 1**: Independent-State Deep Quality Network Learning (isDQN)

Initialize replay memory D to capacity N
Initialize action-value function Q with random weights $\theta$
**For** episode = 1, M **do**
    Initialize sequence $s_1 = \{x_1\}$ and preprocessed sequence $\phi_1 = \phi(s_1)$
    **For** t = 1,T **do**
        Store transition $(\phi_t, a_t, r_t)$ in D
        Every b steps **do**
            Sample random minibatch of transitions $(\phi_j, a_j, r_j)$ from D
            Set $y_j = r_j$
            Perform a gradient descent step on $\ell(y_j, Q(\phi_j, a_j; \theta))$ with respect to the network parameters $\theta$
    **End for**
**End for**

($n_{samples}$=52,801). In order to ensure sufficient training data for each treatment, treatments with <200 samples were also removed from the dataset (and therefore samples to which these treatments were administered, resulting in $n_{samples}$=48,335). Variable preprocessing was performed to achieve a format suitable for one-hot encoding, with numeric variables also treated as categorical. Age was then binned into four categories: 'Under 40' (n=16,165), '40-50' (n=12,794), '50-60' (n=13,699), and 'Above 60' (n=17,952). Finally, variables were filtered to remove those which were reflective of treatment outcome and to drop variables with more than 50 categories such as dates of visit, record ID, and physician comments ($n_{var}$=163). Variables were one-hot encoded for upstream analysis. ($n_{samples}$=48,335, $n_{one\text{-}hot\_encodings}$=866)

*PPI Dose Mapping*

PPI doses were mapped to provide a structured framework for interpreting dose variability between different PPIs. Omeprazole doses of 80mg are considered 'High Dose' whereas doses of 10, 20, and 40mg are considered 'Standard or Low Dose'. For Lansoprazole, doses of 60 mg are considered high and 30 or 15mg are considered standard or low. For Pantoprazole, all doses are considered standard of low (including 20 and 40mg) whereas for Esomeprazole, 40 and 80mg are considered high, with 20 mg considered standard or low. Finally, Rabeprazole doses of 40mg are considered high whereas 20 and 10mg are considered standard or low. Only 501 patients were given a dose deviating from these values, which we consider 'non-traditional' and perhaps mis-entered, omitting this information and grouping with unlisted PPI values. Overall, 30,086 patients were given PPI doses considered 'High Dose', 20,180 were given doses considered 'Standard or Low Dose'. A large fraction of the dataset did not have a specified PPI dose by clinician, and when combined with non-traditional doses 22,047 termed 'Nondescript Dose' were included in the dataset.

*Action Space Definition*

The action space was defined in terms of antibiotic/PPI combination prescribed by the clinician as well as the dose of each medication, number of intakes (for example, twice a day), PPI dose category, and duration it was prescribed. For example, a quadruple bismuth regimen of clarithromycin (dose=500mg, twice daily), amoxicillin (1000mg, twice daily), bismuth salts (120mg, four times daily), and high dose PPI for a 14-day duration would be represented as a single numeric action. The action space was restricted to include only treatments with at least 500 examples in the dataset.

*Independent-State Deep Quality Network Learning (isDQN)*

The analysis method used to train our recommendation system, termed independent-state Deep Quality Network Learning (isDQN) takes the form of a traditional DQN analysis, except that optimizations to network weights do not consider a subsequent state when calculating loss. A deep neural network of four layers was implemented for this analysis, consisting of an input layer ($n_{nodes}$=866, the number of one-hot encoded variables for each patient), two hidden layers ($n_{nodes}$=128 each), and an output layer ($n_{nodes}$=12, the total number of combinations of treatment and duration). Inputs to the optimization step consist of:

- $\phi$, the state space, where $\phi = \{\phi_1, \dots, \phi_{n_{state}}\}$ and represents the set of binary (one-hot) encoded patient variables, and $n_{state}$=1265.
- $A$, the action space, where $A = \{a_1, \dots, a_{n_{action}}\}$ and represents the numerically encoded combination of treatment and durations of antibiotic combinations with durations of 7, 10, and 14 days, and therefore $n_{action}$=12
- $R$, the reward space, where $R = \{-1, +1\}$ and +1 represents successful eradication, whereas -1 represents failed eradication.

Optimization was performed via gradient descent on a smooth L1 loss function, which was chosen due to its lessened sensitivity to outliers and tendency to avoid exploding gradients. For a batch size N, loss is defined as:

$$l(x, y) = mean(L) = mean(\{l_1, \ldots, l_N\}^T) \quad (1)$$

Where:

$$l_n = \begin{cases} \frac{(x_n - y_n)^2}{2*\beta}, & if\ |x_n - y_n| < \beta \\ |x_n - y_n| - \frac{\beta}{2}, & otherwise \end{cases} \quad (2)$$

Where $\beta = 1$, and $N = number\ of\ patients\ per\ batch$, and

$$x_n = R_n \quad (3)$$

Where $R_n$ is the reward observed for patient n,

$$y_n = Q(\phi_n, a_n; \theta) \quad (4)$$

Where $Q(s_n, a_n)$ is the quality score for the state-action (patient-treatment) pair of patient n determined by a forward pass through the neural network and $\theta$ are the neural network parameters.

The isDQN algorithm is modified from the original DQN algorithm detailed in Mnih et al.[33] in **Algorithm 1.**

*Selection of optimal treatment for AI recommendation*

The AI-recommended action for a given patient is defined by:

$$a^*(\phi_n) \leftarrow argmax_a\ Q(\phi_n; \theta)$$

Where the argmax function determines the state-action pair of highest Q-value by considering the quality of all possible actions for a given patient, defined by patient variables $\phi_n$. Therefore, Q-scores of AI actions can be directly compared the Q-scores of Clinical decisions via $Q(\phi_n, a^*(\phi_n); \theta))$ and $Q(\phi_j, a_j; \theta)$, respectively.

*Theoretical AI Clinician Testing*

A synthetic dataset of 10000 patients with 100 binary variable features was generated. To model a dataset with imbalanced classes, patients were split into two groups of 7000 and 3000 patients respectively where binary variables were identical within groups by a random selection of 0's and 1's. Next, noise was added to the dataset by introducing a random selection of a 0 or 1 at random intervals in each patient to a desired level. Datasets with noise levels of 5, 10, 25, 50, 75, and 99%% were generated. Each group of patients was 'treated' half of the time with treatment A and half of the time with treatment B, where treatment A was made 90% effective in group A and 80% effective in group B. Treatment B was made 90% effective in group B and 80% effective in group A. Rewards of +1 were assigned for 'successful' treatments and -1 for failures at the 90% and 80% rate described above. Training was performed on a balanced randomized 90-10 training-testing split, where optimization was carried out using Mean Square Error (MSE) loss and hyperparameters of batch size 1000, learning rate 0.001, deque size of 1000, and steps to optimize of 100. Evaluation of results was performed by counting the number of times a treatment with 90% effectiveness in its respective group was recommended to a patient from this respective group, for example treatment A recommended to a patient from group A.

*Sensitivity Analysis and Hyperparameter Selection*

To determine optimal hyperparameters in the context of a more complex dataset similar to *Hp*-EuReg, 50,000 patients were generated with 10 unique 'types' of 70 patient variables. For each of the ten groups, one treatment was given 90% effectiveness while all others were given 70% effectiveness, with no treatment being 90% effective in any two groups of patients. A hyperparameter grid was tested including: batch size of 1000, 5000, and 10000; a learning rate of $5*10^{-5}$, $10^{-4}$, $5*10^{-4}$, and $10^{-3}$; steps to optimize of 50 and 100; and a deque size of 1000, 5000, and 10000. Scoring of best hyperparameters was based on the fraction of treatments correctly recommended to patient group based on higher efficiency in training data. The hyperparameters chosen for further modeling had the highest percentage of correct recommendations of 99.9%, and consisted of a batch size of 1000, learning rate of $5*10^{-5}$, and deque size of 1000.

*Single Model Examination*

To examine average trends, Q scores were broken down by treatment categories including quadruple bismuth therapies with clarithromycin, amoxicillin, and bismuth salts, quadruple non-bismuth therapies with clarithromycin, amoxicillin, and metronidazole, sequential therapies with clarithromycin, amoxicillin, and tinidazole, triple therapies with clarithromycin and amoxicillin, triple therapies with clarithromycin and metronidazole, and Pylera®. All treatments include a PPI in the formulation. Treatments are also broken down by PPI dose into categories of low/standard and high. The subset of patients recommended a nondescript PPI dose was dropped from this analysis. Finally, Q scores are examined on average by duration, including seven-, ten-, and fourteen-day formulations. Mean and standard deviation for each of these groups was calculated to achieve a comparative ranking of each treatment component.

*Patient Specific Recommendation Analysis*

Ten-fold, fifty repeat cross-validation was chosen to evaluate the quality and consistency of our analysis. In ten-fold cross validation, the dataset is divided into ten random samples or folds. Iteratively each fold is taken to represent a 'testing' subset of the data, whereas the remaining nine are taken as a 'training' subset, therefore applying a 90-10 percent training-testing split. Network training (optimization) is performed via isDQN analysis as described above, where the quality of AI and clinical decision is assessed via Q-score at each patient iteration. During testing, no further optimizations are performed and only the quality of AI and clinical decisions are assessed. This process is repeated fifty times to check for bias in random samples and to evaluate the most common recommendation for each patient after many repeats. The recommended treatments in the testing phase of each model were tabulated for each patient, resulting in 50 recommendations total per patient. The mode treatment for each person was recorded to examine heterogeneity in the treatment recommendation.

*Determining Relationship of Patient Variables to Treatment Recommendation by Random Forest*

RF models are generated for each of four treatment categories: Pylera®, bismuth quadruple (including clarithromycin, amoxicillin, and bismuth salts), bismuth salts (any) – which includes either Pylera® or bismuth quadruple therapy in the formulation described above, and non-bismuth quadruple therapy including clarithromycin amoxicillin, and metronidazole. The RF model is formulated to predict whether a patient will receive a given treatment versus all others. Patient variables are then ranked to determine size of effect in the model on predicting outcome using a metric of mean decrease of impurity (MDI). Full lists of ranked variables and MDI are available in the *Supplement.*

AUTHOR CONTRUBUTIONS

K.H. was responsible for methodological development, simulation design and experiment, evaluation of results and manuscript preparation. K.V., D.V., I.L. and T.F.K. worked on conceptualization, study design, methodology formulation/developments and data analysis. Members of the AIDA consortium contributed to the project design and conceptualisation. O.G. managed data collection, curation, and advising on processing and interpreting results. K.V. supervised method development, simulations, and evaluations. J.S. and I.L. provided critical advice on bioinformatic and machine learning analysis, including evaluation and interpretation of results. J.G. was responsible for overseeing data collection and interpretation of results. T.F.K. and K.V. were responsible for securing funding and overall project management. All authors contributed to the writing and editing of the manuscript and approved the final version.

CONFLICT OF INTEREST STATEMENT

Javier P. Gisbert has served as speaker, consultant, and advisory member for or has received research funding from Mayoly Spindler, Allergan, Diasorin, Richen, Biocodex and Juvisé.

Olga P. Nyssen received research funding from Allergan, Mayoly Spindler, Richen, Biocodex and Juvisé.

Drs Kirill Veselkov, Ivan Laponogov, and Dennis Veselkov are affiliated with Intelligify Ltd, an AI consultancy company, which was not involved in the research, analysis, or interpretation of the results presented in this study.

Tania Fleitas Kanonnikoff discloses advisory roles honoraria from Amgen, AstraZeneca, Beigene, BMS and MSD. Institutional research funding from Gilead. Speaker honoraria from Amgen, Servier, BMS, MSD, Lilly, Roche, Bayer.

The remaining authors declare no conflicts of interest.

ACKNOWLEDGEMENTS

We thank the Spanish Association of Gastroenterology (AEG) for providing the e-CRF service free of charge. Figures 1, 2, and 3A made in Biorender (www.biorender.com)

DATA AVAILABILITY STATEMENT

The data supporting the conclusions of this study are not publicly available, as their content may compromise the privacy of research participants. However, previously published data from the Hp-EuReg study, or de-identified raw data referring to the current study, as well as further information on the methods used to explore the data, may be shared by contacting opn.aegredcap@aegastro.es, with no particular time constraint. Individual participant data will not be shared. Code has been made available at the following public bitbucket repository: https://bitbucket.org/iAnalytica/aiclinician/src/main/.

FUNDING

This research was supported by the AIDA project, funded by UK Research and Innovation (Grant No. 10058099) and the European Union (Grant No. 101095359).

Data for AI/computational developments were provided by the HP-EuReg registry, which was promoted and funded by the European Helicobacter and Microbiota Study Group (EHMSG). Additional support was received from the Spanish Association of Gastroenterology (AEG) and the Centro de Investigación Biomédica en Red de Enfermedades Hepáticas y Digestivas (CIBERehd).

The HP-EuReg was co-funded by the European Union's Horizon Europe programme (Grant Agreement No. 101095359) and supported by UK Research and Innovation (Grant Agreement No. 10058099). It was also co-funded by the European Union's EU4Health programme (Grant Agreement No. 101101252).

The views and opinions expressed are those of the author(s) and do not necessarily reflect those of the European Union or the Health and Digital Executive Agency (HaDEA). Neither the European Union nor the granting authorities bear responsibility for the content.

The HP-EuReg study was additionally funded by Diasorin, Juvisé, and Biocodex. However, these companies had no access to clinical data and were not involved in any stage of the study, including its design, data collection, statistical analysis, or manuscript preparation. We acknowledge their financial support with gratitude.

Supplementary Information 1 **AIDA Consortium Members and Affiliations.**

1. **Tania Fleitas Kanonnikoff, Ana Miralles Marco, Manuel Cabeza-Segura, Elena Jiménez Martí, Josefa Castillo.**

    Instituto Investigación Sanitaria INCLIVA (INCLIVA), Medical Oncology Department, Hospital Clínico Universitario de Valencia; Avda Blasco Ibáñez, 17, 46010, Valencia, Spain.

    Biochemistry and Molecular Biology Department, Universitat de València, Avda Blasco Ibáñez, 15, 46010, Valencia, Spain

2. **Kirill Veselkov**

    Imperial College of London (ICL), Department of Surgery and Cancer, Sir Alexander Fleming Building, South Kensington Campus, SW7 2AZ, London, UK.

3. **Mārcis Leja, Inese Poļaka**

    Institute of Clinical and Preventive Medicine, Faculty of Medicine and Lifesciences, University of Latvia, LV-1586 Riga, Latvia

4. **Fatima Carneiro, Ceu Figueiredo, Rui Ferreira, Rita Barros.**

    i3S - Instituto de Investigação e Inovação em Saúde, Universidade do Porto, Rua Alfredo Allen 208, 4200-135 Porto, Portugal

    Ipatimup - Institute of Molecular Pathology and Immunology of the University of Porto, Porto, Portugal

    Faculty of Medicine of the University of Porto, Portugal

    Department of Pathology, Unidade Local de Saúde São João, Porto, Portugal

5. **Javier P. Gisbert, Olga P.Nyssen**

    Gastroenterology Unit, Hospital Universitario de La Princesa, Instituto de Investigación Sanitaria Princesa (IIS-Princesa), Centro de Investigación Biomédica en Red de Enfermedades Hepáticas y Digestivas (CIBERehd), Universidad Autónoma de Madrid (UAM), Diego de León, 62, 28006, Madrid, Spain

6. **Leticia Moreira, Miriam Cuatrocasas, Gloria Fernandez-Esparrach**

    Department of Gastroenterology, Hospital Clínic de Barcelona; Institut d'Investigacions Biomèdiques August Pi i Sunyer (IDIBAPS); Centro de Investigación Biomédica en Red de Enfermedades Hepáticas y Digestivas (CIBEREHD), Facultat de Medicina i Ciències de la Salud, Universitat de Barcelona (UB), 08036 Barcelona, Spain

7. **Tamara Matysiak-Budnik, Jerome Martin**

    Institut des Maladies de l'Appareil Digestif, Hépato-Gastroentérologie, Hôtel Dieu, Centre Hospitalier Universitaire. 1 Place Alexis Ricordeau, 44093 Nantes, France

8. **Laimas Jonaitis, Juozas Kupčinskas, Paulius Jonaitis**

    Department of Gastroenterology, Lithuanian University of Health Sciences, Eiveniu Street 2, 50161, Kaunas, Lithuania

9. **Mário Dinis-Ribeiro, Miguel Coimbra, Ana Carina Pereira, Filipa Fontes**

    RISE@CI-IPOP (Health Research Network), Portuguese Oncology Institute of Porto (IPO Porto), 4200-072 Porto, Portugal

10. **Manon C.W. Spaander**

    Department of Gastroenterology and Hepatology, Erasmus University Medical Center, Rotterdam, the Netherlands.

11. **Stefano Sedola, Junior Andrea Pescino**

StratejAI, Avenue Louise 209, 1050 Brussels, Belgium

12. **Zorana Maravic, Ana Martins**

    Digestive Cancers Europe, Rue de la Loi 235/27, 1040 Brussels, Belgium

Supplementary Table 1 Random Forest Impotance Scores

**Quadruple Bismuth CABS**

| Feature | Importance (MDI) |
|---|---|
| Region:East | 0.21512057652581568 |
| Is the patient taking Rebamipid?:NA | 0.14666938065279966 |
| Ethnic Background :Caucasian | 0.11571243385360448 |
| Is the patient taking Rebamipid?:No | 0.11415842483321086 |
| Ethnic Background :Other | 0.10565510858570352 |
| Region:South-west | 0.0596739 |
| Is the patient taking pro/pre-biotics? (choice=Yes - Probiotics)Unchecked | 0.04053333 |
| Region:East-centre | 0.025872425210350544 |
| Is the patient taking Rebamipid?:Yes | 0.020654279586257658 |
| Indication :Dyspepsia with normal endoscopy | 0.00990902 |
| Indication :Non Investigated Dyspepsia | 0.00825233 |
| Region:West-centre | 0.00750417 |
| Gastrointestinal symptoms (choice=Heartburn)Unchecked | 0.0071195 |
| Is the patient taking any concurrent medication?:No | 0.00707658 |
| Age calculated bins:Under 40 | 0.00634923 |
| Gastrointestinal symptoms (choice=Dyspepsia)Checked | 0.00628107 |
| Acetylsalicylic acid:Not taking | 0.00579321 |
| Gastrointestinal symptoms (choice=Other)Checked | 0.00533652 |
| Age calculated bins:Above 60 | 0.00502646 |
| Statins:Not taking | 0.00502234 |
| Ethnic Background :Asian | 0.00485868 |
| Statins:NA | 0.00484108 |
| Ethnic Background :NA | 0.0047632 |
| Gender:Female | 0.00447509 |
| Gender:Male | 0.00444433 |
| Age calculated bins:50 - 60 | 0.00418828 |
| Acetylsalicylic acid:NA | 0.00412597 |
| Age calculated bins:40 - 50 | 0.0038354 |
| Proton pump inhibitors:Daily | 0.00368462 |
| Region:North | 0.00368004 |
| Gastrointestinal symptoms (choice=None)Unchecked | 0.00312806 |
| Indication :Gastric Ulcer | 0.00304101 |
| Is the patient taking any concurrent medication?:Yes | 0.00284829 |
| Indication :Duodenal Ulcer | 0.00284411 |
| Indication :Other | 0.00260715 |
| Proton pump inhibitors:NA | 0.00244999 |
| Indication :Preneoplastic lesions (atrophic gastritis or intestinal metaplasia) | 0.0021039 |
| Antibiotic Resistance  (choice=Not performed)Unchecked | 0.00164658 |
| Antibiotic Resistance  (choice=No resistance)Unchecked | 0.00156413 |
| NSAIDs:Not taking | 0.00140277 |
| Acetylsalicylic acid:Daily | 0.00127942 |

| Feature | Importance (MDI) |
| --- | --- |
| NSAIDs:NA | 0.00120232 |
| Is the patient taking any concurrent medication?:NA | 0.00116233 |
| Antibiotic Resistance  (choice=Nitroimidazole)Unchecked | 0.00115652 |
| Proton pump inhibitors:Not taking | 0.00110086 |
| NSAIDs:On demand | 0.00102191 |
| Proton pump inhibitors:On demand | 0.00088744 |
| Indication :Long-term treatment with PPIs | 0.00083609 |
| Region:Other | 0.00079032 |
| Ethnic Background :Black | 0.00067201 |
| Drug allergies:Yes, Specify… | 0.00064732 |
| Drug allergies:No | 0.00063557 |
| Statins:Daily | 0.00062363 |
| Antibiotic Resistance  (choice=Quinolone)Unchecked | 0.00055657 |
| Antibiotic Resistance  (choice=Clarithromycin)Unchecked | 0.00055504 |
| Indication :NSAIDs (or aspirin) treatment | 0.00039661 |
| Indication :First-degree relatives of patients with gastric cancer | 0.00039307 |
| Indication :Unexplained iron deficiency anaemia | 0.00030005 |
| If patient has drug allergies, please specify here: (choice=Penicillin)Unchecked | 0.00020159 |
| Indication :Screening (to prevent gastric cancer) | 0.00016936 |
| Acetylsalicylic acid:On demand | 0.0001595 |
| If patient has drug allergies, please specify here: (choice=Tetracyclines)Unchecked | 0.00014967 |
| Statins:On demand | 0.00013194 |
| Age calculated bins:NA | 0.0001154 |
| NSAIDs:Daily | 0.00010948 |
| Indication :Surgical or endoscopic resection of gastric cancer | 0.00010801 |
| If patient has drug allergies, please specify here: (choice=Bismuth)Unchecked | 9.503927927100457e-05 |
| Indication :MALT lymphoma | 8.254974049668174e-05 |
| Gender:NA | 7.901150725401054e-05 |
| If patient has drug allergies, please specify here: (choice=Fluoroquinolones)Unchecked | 4.39745706586533640e-05 |
| Indication :NA | 3.536997797636809e-05 |
| If patient has drug allergies, please specify here: (choice=Macrolides)Unchecked | 2.03242946705912760e-05 |
| Indication :Idiopathic thrombocytopenic purpura | 1.5817078642395863e-05 |
| Indication :Vitamin B12 deficiency | 1.00337646161248440e-05 |
| Drug allergies:NA | 1.4245047730028902e-06 |
| Antibiotic Resistance  (choice=Amoxicillin)Unchecked | 1.14065944495706460e-06 |
| Antibiotic Resistance  (choice=Tetracycline)Unchecked | 2.9705916304402096e-07 |

**Pylera**

| Feature | Importance (MDI) |
| --- | --- |
| Statins:NA | 0.12708558 |
| Region:South-west | 0.09002151 |
| NSAIDs:NA | 0.07225203 |
| Acetylsalicylic acid:Not taking | 0.06950085 |
| Is the patient taking any concurrent medication?:Yes | 0.06393786 |

| Feature | Value |
|---|---|
| Acetylsalicylic acid:NA | 0.0609539 |
| Proton pump inhibitors:NA | 0.0512968 |
| Is the patient taking any concurrent medication?:No | 0.03937626 |
| Statins:Not taking | 0.03605225 |
| Region:East-centre | 0.03481504 |
| Proton pump inhibitors:Daily | 0.02419517 |
| NSAIDs:Not taking | 0.01851988 |
| NSAIDs:On demand | 0.01678415 |
| Is the patient taking pro/pre-biotics? (choice=Yes - Probiotics)Unchecked | 0.01597801 |
| Gastrointestinal symptoms (choice=Heartburn)Unchecked | 0.01489391 |
| Proton pump inhibitors:Not taking | 0.01347018 |
| Indication :Dyspepsia with normal endoscopy | 0.0133038 |
| Age calculated bins:40 - 50 | 0.01310273 |
| Gastrointestinal symptoms (choice=Dyspepsia)Checked | 0.01241311 |
| Gender:Female | 0.01209852 |
| Gender:Male | 0.01209048 |
| Region:East | 0.01155117 |
| Statins:Daily | 0.01118411 |
| Age calculated bins:Above 60 | 0.01101182 |
| Gastrointestinal symptoms (choice=Other)Checked | 0.01084361 |
| Age calculated bins:50 - 60 | 0.01024544 |
| Indication :Non Investigated Dyspepsia | 0.01023239 |
| Age calculated bins:Under 40 | 0.00956397 |
| Indication :Duodenal Ulcer | 0.00874273 |
| Proton pump inhibitors:On demand | 0.00827331 |
| Acetylsalicylic acid:Daily | 0.00754108 |
| Indication :Gastric Ulcer | 0.0060181 |
| Region:West-centre | 0.00598397 |
| Ethnic Background :Caucasian | 0.00595107 |
| Indication :Other | 0.00579154 |
| Gastrointestinal symptoms (choice=None)Unchecked | 0.00515169 |
| Indication :Unexplained iron deficiency anaemia | 0.00497868 |
| Drug allergies:No | 0.00458752 |
| Antibiotic Resistance  (choice=No resistance)Unchecked | 0.00405765 |
| Drug allergies:Yes, Specify... | 0.00404202 |
| If patient has drug allergies, please specify here: (choice=Penicillin)Unchecked | 0.00372875 |
| Indication :Preneoplastic lesions (atrophic gastritis or intestinal metaplasia) | 0.00363117 |
| Is the patient taking Rebamipid?:NA | 0.00362709 |
| Region:Other | 0.00351634 |
| Ethnic Background :Other | 0.00334927 |
| Indication :First-degree relatives of patients with gastric cancer | 0.0029259 |
| NSAIDs:Daily | 0.00277664 |
| Antibiotic Resistance  (choice=Nitroimidazole)Unchecked | 0.00195794 |
| Indication :Vitamin B12 deficiency | 0.00177304 |

| Feature | Importance |
|---|---|
| Region:North | 0.00170139 |
| Ethnic Background :NA | 0.00146731 |
| Antibiotic Resistance  (choice=Not performed)Unchecked | 0.0013512 |
| Antibiotic Resistance  (choice=Quinolone)Unchecked | 0.00134378 |
| Is the patient taking Rebamipid?:No | 0.00125736 |
| If patient has drug allergies, please specify here: (choice=Fluoroquinolones)Unchecked | 0.00101212 |
| Indication :Long-term treatment with PPIs | 0.00086438 |
| Acetylsalicylic acid:On demand | 0.00078681 |
| Is the patient taking any concurrent medication?:NA | 0.00076703 |
| Antibiotic Resistance  (choice=Clarithromycin)Unchecked | 0.00069331 |
| Ethnic Background :Black | 0.00049886 |
| Statins:On demand | 0.00049043 |
| Ethnic Background :Asian | 0.00041168 |
| Indication :Screening (to prevent gastric cancer) | 0.00031407 |
| Indication :NSAIDs (or aspirin) treatment | 0.00026672 |
| If patient has drug allergies, please specify here: (choice=Macrolides)Unchecked | 0.00023805 |
| Indication :Surgical or endoscopic resection of gastric cancer | 0.00023751 |
| Age calculated bins:NA | 0.00020621 |
| Antibiotic Resistance  (choice=Tetracycline)Unchecked | 0.00017283 |
| Is the patient taking Rebamipid?:Yes | 0.00017279 |
| If patient has drug allergies, please specify here: (choice=Tetracyclines)Unchecked | 0.00016325 |
| Antibiotic Resistance  (choice=Amoxicillin)Unchecked | 0.00013317 |
| Indication :MALT lymphoma | 0.00012279 |
| Indication :Idiopathic thrombocytopenic purpura | 8.76E-05 |
| Drug allergies:NA | 5.28E-05 |
| Indication :NA | 3.33E-06 |
| Gender:NA | 1.71E-06 |
| If patient has drug allergies, please specify here: (choice=Bismuth)Unchecked | 1.47E-06 |

**Bismuth (Pylera or CABS)**

| Feature | Importance (MDI) |
|---|---|
| Region:East | 0.06234324 |
| Is the patient taking any concurrent medication?:Yes | 0.03658758 |
| Statins:NA | 0.03591559 |
| Gastrointestinal symptoms (choice=Heartburn)Unchecked | 0.0354624 |
| Proton pump inhibitors:NA | 0.03471757 |
| NSAIDs:NA | 0.033862089905877094 |
| Acetylsalicylic acid:NA | 0.032740971660101315 |
| Is the patient taking pro/pre-biotics? (choice=Yes - Probiotics)Unchecked | 0.03225501 |
| Region:South-west | 0.0319111804506280415 |
| Region:East-centre | 0.03156761 |
| Gastrointestinal symptoms (choice=Dyspepsia)Checked | 0.03014129 |
| Is the patient taking Rebamipid?:NA | 0.028900584649131176 |
| Is the patient taking Rebamipid?:No | 0.028562949492584328 |

| Feature | Value |
|---|---|
| Gastrointestinal symptoms (choice=Other)Checked | 0.025369085837009822 |
| Ethnic Background :Caucasian | 0.024246715326253977 |
| Ethnic Background :Other | 0.023745894677667415 |
| Age calculated bins:40 - 50 | 0.023367785163230056 |
| Statins:Not taking | 0.02288587 |
| Indication :Dyspepsia with normal endoscopy | 0.0207942567437226330 |
| Age calculated bins:50 - 60 | 0.02028303 |
| Acetylsalicylic acid:Not taking | 0.02017641 |
| Age calculated bins:Above 60 | 0.01986522 |
| Gender:Male | 0.0197509691679771460 |
| NSAIDs:Not taking | 0.018919330647481084 |
| Gender:Female | 0.01878453 |
| Age calculated bins:Under 40 | 0.01766835 |
| Indication :Non Investigated Dyspepsia | 0.01715366 |
| Is the patient taking any concurrent medication?:No | 0.015509238902882175 |
| Indication :Duodenal Ulcer | 0.015332901767151082 |
| Indication :Gastric Ulcer | 0.01266041 |
| Gastrointestinal symptoms (choice=None)Unchecked | 0.012546879459148305 |
| Is the patient taking any concurrent medication?:NA | 0.012239642908180595 |
| Indication :Other | 0.011674513776715793 |
| Statins:Daily | 0.010944927720347703 |
| Region:West-centre | 0.010710464173400698 |
| Proton pump inhibitors:Not taking | 0.0105002959192110558 |
| Proton pump inhibitors:Daily | 0.010432647643548834 |
| Antibiotic Resistance  (choice=No resistance)Unchecked | 0.010205511880952603 |
| Antibiotic Resistance  (choice=Not performed)Unchecked | 0.00982732 |
| Antibiotic Resistance  (choice=Nitroimidazole)Unchecked | 0.00845521 |
| Region:North | 0.00740857 |
| NSAIDs:On demand | 0.00699954 |
| Indication :Preneoplastic lesions (atrophic gastritis or intestinal metaplasia) | 0.00657606 |
| Indication :Unexplained iron deficiency anaemia | 0.00618617 |
| Proton pump inhibitors:On demand | 0.00614096 |
| Antibiotic Resistance  (choice=Clarithromycin)Unchecked | 0.00579314 |
| Region:Other | 0.0057385 |
| Drug allergies:Yes, Specify... | 0.00561695 |
| Antibiotic Resistance  (choice=Quinolone)Unchecked | 0.00556766 |
| Drug allergies:No | 0.0046274 |
| Ethnic Background :NA | 0.00455807 |
| Acetylsalicylic acid:Daily | 0.00455429 |
| Indication :First-degree relatives of patients with gastric cancer | 0.00449764 |
| If patient has drug allergies, please specify here: (choice=Penicillin)Unchecked | 0.00339467 |
| Ethnic Background :Asian | 0.0030308 |
| Is the patient taking Rebamipid?:Yes | 0.00278528 |
| Indication :Long-term treatment with PPIs | 0.00267369 |

| Feature | Importance (MDI) |
| --- | --- |
| Ethnic Background :Black | 0.00261955 |
| NSAIDs:Daily | 0.00256022 |
| Indication :Vitamin B12 deficiency | 0.00206522 |
| Indication :Screening (to prevent gastric cancer) | 0.00090928 |
| Indication :NSAIDs (or aspirin) treatment | 0.00074092 |
| Antibiotic Resistance  (choice=Amoxicillin)Unchecked | 0.00065407 |
| Acetylsalicylic acid:On demand | 0.00065342 |
| Indication :MALT lymphoma | 0.00061846 |
| Statins:On demand | 0.0005772 |
| Indication :NA | 0.00051778 |
| Indication :Idiopathic thrombocytopenic purpura | 0.00050851 |
| Indication :Surgical or endoscopic resection of gastric cancer | 0.00048599 |
| If patient has drug allergies, please specify here: (choice=Tetracyclines)Unchecked | 0.00046533 |
| Antibiotic Resistance  (choice=Tetracycline)Unchecked | 0.00040977 |
| If patient has drug allergies, please specify here: (choice=Fluoroquinolones)Unchecked | 0.00026808 |
| Drug allergies:NA | 0.00023377 |
| Age calculated bins:NA | 0.00017683 |
| Gender:NA | 0.00016972 |
| If patient has drug allergies, please specify here: (choice=Macrolides)Unchecked | 0.00015584 |
| If patient has drug allergies, please specify here: (choice=Bismuth)Unchecked | 3.8213072007855646e-05 |

**Quadruple Non-Bismuth CAM**

| Feature | Importance (MDI) |
| --- | --- |
| Is the patient taking any concurrent medication?:Yes | 0.06296587 |
| Region:East | 0.05003785 |
| Statins:NA | 0.04356555 |
| NSAIDs:NA | 0.04276232 |
| Proton pump inhibitors:NA | 0.04092454 |
| Is the patient taking pro/pre-biotics? (choice=Yes - Probiotics)Unchecked | 0.03881023 |
| Acetylsalicylic acid:NA | 0.03603606 |
| Gastrointestinal symptoms (choice=Heartburn)Unchecked | 0.03344275 |
| Gastrointestinal symptoms (choice=Other)Checked | 0.02919808 |
| Gastrointestinal symptoms (choice=Dyspepsia)Checked | 0.02885517 |
| Region:South-west | 0.02882487 |
| Region:West-centre | 0.02688091 |
| Is the patient taking Rebamipid?:NA | 0.02484127 |
| Indication :Dyspepsia with normal endoscopy | 0.02446547 |
| Region:East-centre | 0.0244229 |
| Age calculated bins:40 - 50 | 0.02267692 |
| NSAIDs:Not taking | 0.02219208 |
| Statins:Not taking | 0.02207089 |
| Is the patient taking Rebamipid?:No | 0.02152578 |
| Ethnic Background :Caucasian | 0.02107499 |
| Indication :Non Investigated Dyspepsia | 0.0195291 |

| Feature | Value |
|---|---|
| Age calculated bins:Above 60 | 0.01819075 |
| Ethnic Background :Other | 0.01799324 |
| Acetylsalicylic acid:Not taking | 0.01770569 |
| Age calculated bins:50 - 60 | 0.01764295 |
| Gender:Male | 0.01756347 |
| Age calculated bins:Under 40 | 0.01745011 |
| Gender:Female | 0.01706945 |
| Is the patient taking any concurrent medication?:No | 0.01660289 |
| Is the patient taking any concurrent medication?:NA | 0.01403877 |
| Proton pump inhibitors:Not taking | 0.01369719 |
| Indication :Duodenal Ulcer | 0.01301609 |
| Gastrointestinal symptoms (choice=None)Unchecked | 0.01179174 |
| Indication :Other | 0.01080212 |
| Indication :Gastric Ulcer | 0.01007286 |
| Antibiotic Resistance  (choice=No resistance)Unchecked | 0.00918605 |
| Proton pump inhibitors:Daily | 0.00902675 |
| Statins:Daily | 0.0084092 |
| Antibiotic Resistance  (choice=Not performed)Unchecked | 0.00787298 |
| Region:North | 0.00686815 |
| Antibiotic Resistance  (choice=Nitroimidazole)Unchecked | 0.00629392 |
| Indication :Preneoplastic lesions (atrophic gastritis or intestinal metaplasia) | 0.0060185 |
| Indication :Unexplained iron deficiency anaemia | 0.00518574 |
| Region:Other | 0.00503876 |
| Drug allergies:Yes, Specify... | 0.00496492 |
| Ethnic Background :NA | 0.00485772 |
| Drug allergies:No | 0.00477151 |
| NSAIDs:On demand | 0.00450748 |
| Proton pump inhibitors:On demand | 0.00432997 |
| Antibiotic Resistance  (choice=Quinolone)Unchecked | 0.00395791 |
| Antibiotic Resistance  (choice=Clarithromycin)Unchecked | 0.00391909 |
| Indication :First-degree relatives of patients with gastric cancer | 0.00343452 |
| If patient has drug allergies, please specify here: (choice=Penicillin)Unchecked | 0.00293327 |
| Ethnic Background :Asian | 0.00271853 |
| Acetylsalicylic acid:Daily | 0.0023833 |
| Ethnic Background :Black | 0.0023439 |
| Indication :Vitamin B12 deficiency | 0.00178798 |
| Indication :Long-term treatment with PPIs | 0.00166823 |
| Is the patient taking Rebamipid?:Yes | 0.00138596 |
| Indication :Screening (to prevent gastric cancer) | 0.00103332 |
| NSAIDs:Daily | 0.00092316 |
| Indication :NA | 0.00077383 |
| Indication :Idiopathic thrombocytopenic purpura | 0.00075639 |
| Indication :MALT lymphoma | 0.00070713 |
| Indication :Surgical or endoscopic resection of gastric cancer | 0.00062454 |
| Indication :NSAIDs (or aspirin) treatment | 0.00056807 |
| Drug allergies:NA | 0.00044561 |
| If patient has drug allergies, please specify here: (choice=Tetracyclines)Unchecked | 0.00022311 |
| If patient has drug allergies, please specify here: (choice=Fluoroquinolones)Unchecked | 0.00020431 |
| Antibiotic Resistance  (choice=Tetracycline)Unchecked | 0.00018722 |
| If patient has drug allergies, please specify here: (choice=Macrolides)Unchecked | 0.00018622 |
| Age calculated bins:NA | 0.00018187 |
| Statins:On demand | 0.00016627 |
| Gender:NA | 0.00014322 |
| Acetylsalicylic acid:On demand | 0.00010246 |
| Antibiotic Resistance  (choice=Amoxicillin)Unchecked | 8.42E-05 |
| If patient has drug allergies, please specify here: (choice=Bismuth)Unchecked | 8.38E-05 |